\documentclass[a4paper,10pt,fleqn]{amsart}
\usepackage{epsfig}
\usepackage{graphics,graphicx}
\usepackage{amsmath}
\usepackage{amssymb}
\usepackage{hyperref}
\usepackage{lipsum}

\chardef\at=`\@
\DeclareRobustCommand{\qed}{%
  \ifmmode 
  \else \leavevmode\unskip\penalty9999 \hbox{}\nobreak\hfill
  \fi
\quad\hbox{\qedsymbol}}
\newcommand{\mathbold}[1]{\mbox{\boldmath $#1$}}
\newcommand{\rz}{\mathbb R}

\newcommand{\pr}{\mathbold P}
\newcommand{\ep}{\mathbb P}

\newcommand{\ex}{\mathbold E}
\newcommand{\ev}{\mathbold V}

\begin{document}






\begin{center}
{\large On $p$-values\\}
\quad\\
Laurie Davies\\
Faculty of Mathematics\\
University of Duisburg-Essen, 45117 Essen, Germany\\
email:laurie.davies@uni-due.de
\end{center}
\quad\\


\section{The use and abuse of $p$-values}
All branches of knowledge which require the analysis of data make use
of $p$-values. Unfortunately in  many cases `make use of' could be
replaced  by `abuse', the many reports of widespread abuse are
convincing. In response The American Statistician 
published a statement on $p$-values by the American Statistical
Association together with supplementary material consisting of
statements by several statisticians and philosophers (\cite{WASLAZ16}). 

The most detailed of the supplementary material is
\cite{GREetal16}. The authors point out that there are many ways in
which any usefulness of a $p$-value can be invalidated. One example is
to perform several experiments and report only the one with the
smallest $p$-value. Problems of this nature will not be discussed
here. It will be assumed that the experiment is so to speak clean and
the data are so to speak high quality. 

\section{Probability models and approximation}
\subsection{Semantics}
There are two meaning of the word `model' is statistics. The first
meaning refers to a parametric family of distributions. Thus the
normal model is the family of all normal distributions. This meaning
of the word `model' is common in much of statistics, in particular
in  Bayesian statistics where such models are the objects of study. 

The second meaning is that of a single probability measure. In this
sense of the word the $N(0,1)$ probability measure is one model, the
$N(0,2)$ probability measure is another model. This is the sense in
which the word will be used in this paper. Models in 
this sense are the atoms so to speak of probability theory and hence
the basic objects of stochastic modelling.  The meaning of the word
`model' in the first sense is a parametric family of models in the
second sense.

\subsection{Approximate models}
The authors of \cite{GREetal16} state 
\begin{equation}
  \tag{A}\label{eq:A}
  \parbox{\dimexpr\linewidth-6em}{%
    \strut
.. the distance between the data and the model prediction is 
measured using a {\it test statistic} ..
    \strut 
  } 
\end{equation} 

and 
\begin{equation} 
  \tag{B}\label{eq:B} 
  \parbox{\dimexpr\linewidth-6em}{%
In logical terms, the $p$-value tests {\it all} the assumptions
about how the data were generated (the entire model) ... 
    \strut
  }
\end{equation} 
Although it is never precisely stated it seems that the word `model'
in the above quotations is meant in the first sense, a parametric family 
of distributions. Whatever the meaning of the word `model' the meaning
of the quotations taken together is clear. The distance between the 
data and the model is based on a test statistic and the corresponding 
$p$-value measures this distance in a particular manner. The phrase 
`In logical terms' in the quotation~(\ref{eq:B}) suggests that in
practice this is not so. Indeed in practice the parametric model is accepted 
and the $p$-value is based on a particular hypothesis $H_0: \mu=\mu_0$
using a statistic especially designed for testing this null 
hypothesis, for example a $t$-test. Such a single statistic cannot  
possibly test `{\it all} the assumptions about how the data were  
generated (the entire model)'.  

A similar attitude is taken in \cite{BIRN62}: consideration is
restricted to 
\begin{equation}
  \tag{C}\label{eq:C}
  \parbox{\dimexpr\linewidth-6em}{%
    \strut
models whose adequacy is postulated and not in
question ... the adequacy of any such model is typically
supported, more or less adequately, by a complex informal synthesis of
previous experimental evidence of various kinds and theoretical
considerations concerning both subject-matter and experimental
techniques. 
\strut
  }
\end{equation}

In contrast to the word `adequacy' being applied to a family of
probability measures it will here be applied to individual probability
measures. Thus the $N(0,1)$ distribution may or may not be adequate
for a given data set.  The only sense I can make of applying the word
`adequate' to a parametric family of probability measures is that
there are values of the parameter for which the individual
distributions specified by these parameter values are consistent with
the data. In general this will be a strict subset of the parameter
space: it is difficult to imagine a data set for which the $N(0,1)$
model and the $N(100,10^{-6})$ model are both adequate or consistent
with the data.

The two different meanings of the word `model' are not just a question
of notation or definition. They reflect two different approaches to
statistics. This may be seen in \cite{BIRN62} where a parametric
family of probability measures has to be adequate without specifying
the adequacy of any individual measure. This is necessary as the
Likelihood Principle requires the proportionality of two different
densities for all values of the parameter and not just for the
adequate ones.  A similar problem occurs when testing hypotheses. The
parametric model is declared adequate without specifying the adequate
values of the parameter. A hypothesis $H_0: \mu=\mu_0$ is then tested to
see whether $\mu=\mu_0$ is consistent with the data. It only makes sense
to do this if the adequate parameter values have not been specified
when declaring the whole family to be adequate as otherwise the test
would be superfluous.

More generally a common approach is two perform a statistical analysis
in two stages. In the first stage one or several parametric models will
be investigated for adequacy, for example by using a goodness-of-fit
test. Once an adequate model has been found it is made the basis of
the second stage where it is treated as if it were true. Treating it
as true means among other things ignoring the first stage. If indeed
the model is now treated as if true then how we arrived at this truth is
irrelevant. The following quotation is taken from the Chapter 5  of
Huber \cite{HUB11} entitled `Approximate Models':
\begin{equation}
  \tag{D}\label{eq:D}
  \parbox{\dimexpr\linewidth-6em}{
    \strut
In the opposite case, if a goodness-of-fit test does not reject a
model, statisticians have become accustomed to acting as if it were
true. Of course this is logically inadmissible, even more so if with
McCullagh and Nelder one believes that all models are wrong {\it a
  priori}.
\quad\\
Moreover, treating a model that has not been rejected as correct can
be misleading and dangerous. Perhaps this is the main lesson we have
learned from robustness.
\strut
  }
\end{equation}

In \cite{DAV14} models are consistently treated as approximations. The basic
idea is that a model $P$  is an adequate approximation to a data
set ${\mathbold x}_n=(x_1,\ldots,x_n)$  if typical data ${\mathbold
  X}_n(P)$ generated 
under $P$ look like ${\mathbold x}_n$. Data are generated under single
probability distributions $P$ and not under a family of such, that is,
a model in the first sense of the word. This is the reason why single
probability distributions are the basic objects of study and not
families of distributions.

The definition of `look like' will depend on the nature of the data
being analysed and the model. As an example suppose that the model is
that of i.i.d. $N(0,1)$ random variables. Then `look like' can be
based on the mean, the variance, the extreme values and the distance of the
empirical distribution function from that of the standard $N(0,1)$
distribution function. This will be done explicitly in
Section~\ref{sec:gauss} below. It is worth noting that the concept of
adequacy is defined in terms of several statistics and not just
one. This is in contrast to the quotation at the beginning of
Section~3.2 where it is based on a single statistic.

The approach described in \cite{DAV14} can be read as
an attempt to replace a two stage methodology, EDA followed by
formal inference, by a single stage methodology whereby all
tests become misspecification tests from without, or from a
distance. It is an instance of `distanced rationality' due to
D.~W.~M\"uller (see \cite{MUELL74}). Here my translation
\begin{equation}
  \tag{E}\label{eq:E}
  \parbox{\dimexpr\linewidth-6em}{%
    \strut
... distanced rationality. By this we mean an attitude to the given,
  which is not governed by any possible or imputed immanent laws but
  which confronts it with draft constructs of the mind in the form of
    \strut
  }
\end{equation}
\begin{equation*}
   \parbox{\dimexpr\linewidth-6em}{%
    \strut
  models, hypotheses, working hypotheses, definitions,
  conclusions, alternatives, analogies, so to speak from a distance, in the manner
  of partial, provisional, approximate knowledge.
    \strut
  }
\end{equation*}

\subsection{`Adequate' parametric families}
Although the quotation~\ref{eq:C} does not make it explicit it is
clear from \cite{BIRN62}  that Birnbaum is referring to families of
parametric models. Thus the Poisson family may be declared adequate
without specifying any particular value of $\lambda$ which is
consistent with the data.  As an example suppose the parametric family
is the Poisson family and that the chi-squared goodness-of-fit is used
to test adequacy. The test is typically based on some variant of the test
statistic 
\begin{equation} \label{equ:chi_pois_1}
  \sum_{j=0}^k\frac{({\hat p}_j-p_j({\bar x}_n))^2}{p_j({\bar x}_n)}
\end{equation}
where the ${\hat p}_j$ are the empirical frequencies, ${\bar x}_n$ is
the mean of the data and $p_j(\lambda)=\lambda^j\exp(-\lambda)/j!$. If
the value of the test statistic (\ref{equ:chi_pois_1})  lies below a
certain level then the Poisson model is declared adequate. Note that
 (\ref{equ:chi_pois_1}) does not specify any individual parameter values. 

Given adequacy in this sense the whole parametric family is then 
transported to the second stage of formal inference in spite of the
fact that the overwhelming majority of individual models will not be
consistent with the data. For Birnbaum's argument to work this is
essential: `two likelihood functions, $f(x,\theta)$ and $g(y,\theta)$
are called the same if they are proportional, that is if there exists
a positive constant $c$ such that $f(x,\theta)=g(y,\theta)$ for all $\theta$'.

In the second sense of the word `model', an individual probability
distribution, the goodness-of-fit procedure takes on a different
form. In the concrete case of the Poisson family a given Poisson
distribution  say  ${\mathcal P}_{\lambda}$ with $\lambda=2$ can be
tested for adequacy using 
\begin{equation} \label{equ:chi_pois_2}
  \sum_{j=0}^k\frac{({\hat p}_j-p_j(2))^2}{p_j(2)}\,.
\end{equation}
The set of $\lambda$ values for which the test statistic lies below a
critical level specifies those $\lambda$ values, if any, which are
consistent with the data. This will not be the set of all possible
values.

If one interprets the concept of adequacy for models in the first
sense using the second sense it can only mean that there are some
parameter values $\theta$ for which the single model $P_{\theta}$ is
consistent with the data. The likelihood principle is based on not
specifying which values of $\theta$ these are.

\subsection{Approximation regions: an example}\label{sec:gauss}
A model $P$ is an adequate approximation to data ${\mathbold x}_n$ if
typical data sets generated under $P$ look like ${\mathbold x}_n$. To
make this susceptible to mathematical analysis the term `look like'
must be expressible in numerical quantities. This may not always be
possible or easy. An animal may be easily recognizable as a dog but it
it not easy to give this a mathematical expression. If a model is
required which gives data sets looking like the daily returns of the
Standard and Poor's 500 index it is not clear how `look like'
can be defined. In the following it will be assumed that `look like'
has a precise mathematical expression. 

The following is taken from \cite{DAV14}. Given a
probability measure $P$ a sample of size $n$  
generated under $P$ will be denoted by ${\mathbold
X}_n(P)=(X_1(P),\ldots,X_n(P))$. Given further a family ${\mathcal P}$
of probability measures and a number $\alpha, 0 <\alpha\le 1,$ an
$\alpha$ approximation region for the data ${\mathbold x}_n$ is defined by
\begin{equation}  \label{equ:approx_reg_1}
{\mathcal A}({\mathbold x}_n,\alpha,{\mathcal P})=\{P \in{\mathcal P}:{\mathbold x}_n
\in E_n(P)\}
\end{equation} 
where for each $P\in{\mathcal P}$ $E_n(P)$ denotes a subset of $\rz^n$
such that
\begin{equation}  \label{equ:approx_reg_2}
\pr({\mathbold X}_n(P) \in E_n(P))=\alpha \,.
\end{equation} 

The choice of the $E_n(P)$ depends on the situation and has in general
to be augmented by some form of regularization, for example: minimum
Fisher models, number of local extremes, convexity constraints. These
and further examples are to be found in \cite{DAV14}.

The definition (\ref{equ:approx_reg_1}) makes no assumption that
 the data ${\mathbold x}_n$ were generated under some model $P_0\in
 {\mathcal P}$. The interpretation is that ${\mathcal A}({\mathbold
   x}_n,\alpha,{\mathcal  P})$ specifies those models $P$ for which
${\mathbold x}_n$ `looks like' a `typical sample' ${\mathbold X}_n(P)$
generated under $P$: typical samples ${\mathbold X}_n(P)$ lie in
$E_n(P)$ so that points ${\mathbold x}_n \in E_n(P)$ look like typical
samples ${\mathbold X}_n(P)$.

As an example suppose ${\mathcal P}$ is the family of normal
distributions ${\mathcal N}=\{(\mu,\sigma):N(\mu,\sigma^2)\}$. An
approximation region can be based on the mean, the variance, outliers
and the distance of the empirical measure to the model $N(\mu,\sigma^2)$ as measured by
the Kuiper metric. More precisely put  ${\mathbold y}_n=({\mathbold
  x}_n-\mu)/\sigma$ and 
\begin{equation} \label{equ:gauss_func_1}
\left\{
\begin{array}{ll}
T_1({\mathbold y}_n)=\sqrt{n}\,\vert \text{mean}({\mathbold y}_n)\vert,&T_2({\mathbold
    y}_n)=\sum_{i=1}^ny_i^2,\\
T_3({\mathbold y}_n)=\max_i \vert y_i\vert,&T_4({\mathbold
    y}_n)=d_{\text{ku}}(\ep({\mathbold y}_n),N(0,1)),\\
\end{array}
\right.
\end{equation}
where $\ep({\mathbold y}_n)$ is the empirical measure based on
${\mathbold y}_n$. Given ${\tilde \alpha}$ one can determine quantiles
$q_1({\tilde \alpha}),q_{21}({\tilde \alpha}),q_{22}({\tilde
  \alpha}),q_3({\tilde \alpha}),q_4({\tilde \alpha})$ such that
\begin{equation} \label{equ:approx_inequ_1}
\left\{
\begin{array}{ll}
\pr(T_1({\mathbold Y}_n)\le q_1({\tilde \alpha}))={\tilde \alpha},& 
\pr(q_{21}({\tilde \alpha})\le T_2({\mathbold
    Y}_n)\le q_{22}({\tilde \alpha}))={\tilde \alpha},\\
\pr(T_3({\mathbold Y}_n)\le q_3({\tilde \alpha}))={\tilde \alpha},&
\pr(T_4({\mathbold Y}_n)\le q_4({\tilde \alpha}))={\tilde \alpha}.\\
\end{array}
\right.
\end{equation}
where ${\mathbold Y}_n$ are i.i.d. $N(0,1)$. The approximation region
is then defined by
\begin{eqnarray}  
\hspace*{0.5cm}{\mathcal A}({\mathbold x}_n,\alpha,\rz\times \rz_+)&=&\{(\mu,\sigma):
T_1({\mathbold y}_n)\le q_1({\tilde \alpha}), q_{21}({\tilde \alpha})\le T_2({\mathbold
    y}_n)\le q_{22}({\tilde \alpha}),  \label{equ:approx_reg_3}\\
&&\quad T_3({\mathbold y}_n)\le q_3({\tilde \alpha}),T_4({\mathbold
  y}_n)\le q_4({\tilde \alpha}),\,\, {\mathbold y}_n=({\mathbold x}_n-\mu)/\sigma\}\nonumber 
\end{eqnarray} 
where ${\tilde \alpha}$ is adjusted so that the region is indeed an
$\alpha$-approximation region. A reasonable starting value for
${\tilde \alpha}$ is $(3+\alpha)/4$. This will lead to an effective
value $\alpha^*>\alpha$ of $\alpha$ which can be determined by
simulations. A better approximation can now be obtained by putting
${\tilde \alpha}=(3+2\alpha-\alpha^*)/4$. For a normal sample of size
$n=50$ and $\alpha=0.9$ this leads to ${\tilde \alpha}\approx 0.97$
compared with the starting value of 0.975.

\begin{figure}[hb]
\begin{center}
\includegraphics[width=4cm,height=12cm,angle=-90]{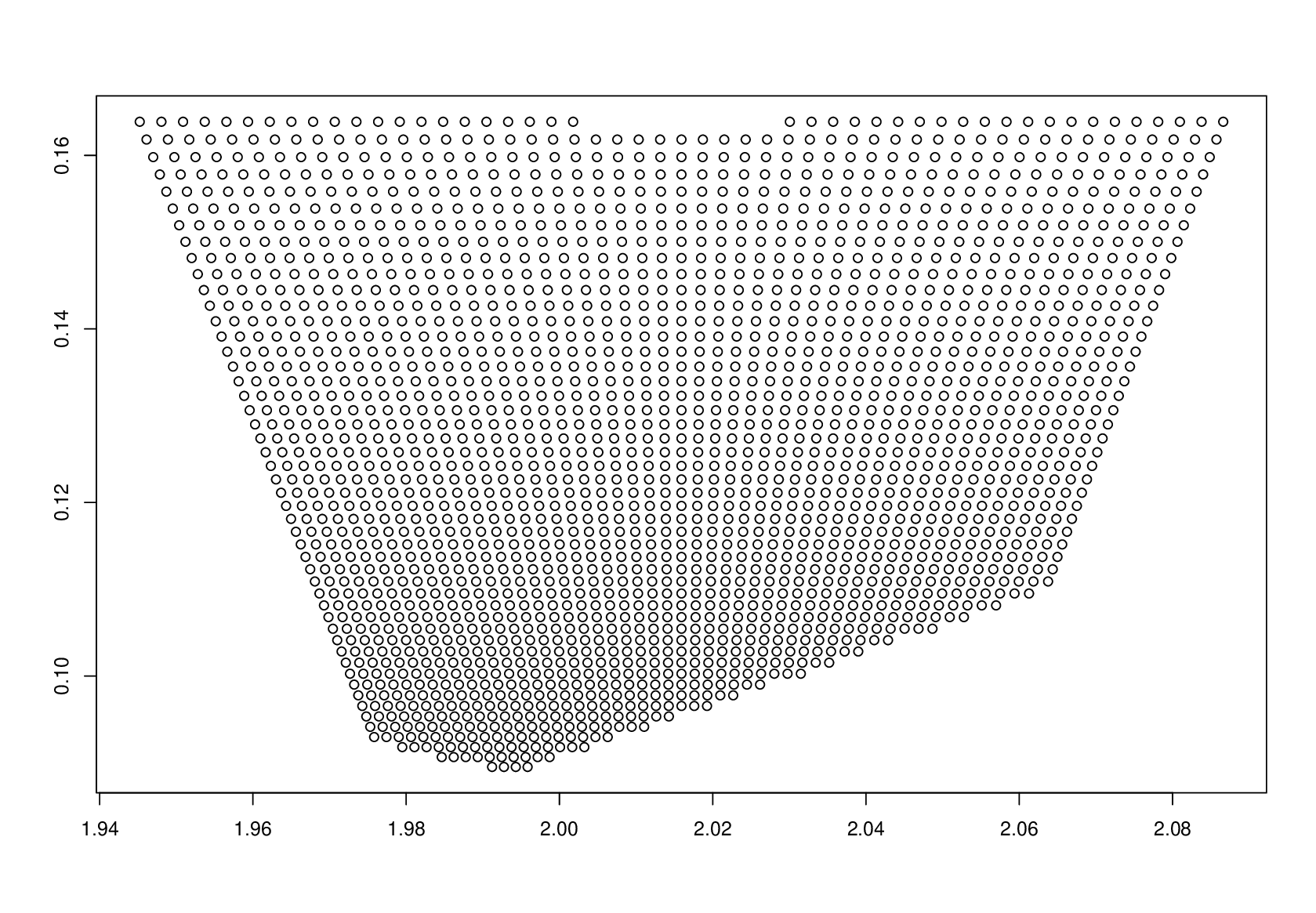}
\end{center}
\caption{The approximation region ${\mathcal A}({\mathbold
  x}_n,0.9,\rz\times \rz_+)$ for the data (\ref{equ:copper_data}). \label{fig:copper_approx}} 
\end{figure}

The following data give the quantity of copper in milligrams per litre
in a sample of drinking water (\cite{DAV14}): 
\begin{eqnarray} \label{equ:copper_data}
&&2.16,2.21,2.15,2.05,2.06,2.04,1.90,2.03,2.06,2.02,2.06,1.92,2.08,\\
&&2.05,1.88,1.99,2.01,1.86,1.70,1.88,1.99,1.93,2.20,2.02,1.92,2.13,2.13.\nonumber
\end{eqnarray}
The 0.9 approximation region ${\mathcal A}({\mathbold
  x}_n,0.9,\rz\times \rz_+)$ this data set is shown in
Figure~\ref{fig:copper_approx}.

An approximation region for $\mu$ alone
can be obtained by projecting ${\mathcal A}({\mathbold
  x}_n,\alpha,{\mathcal N})$ onto the $\mu$-axis:
\begin{equation} \label{equ:mu_approx_1}
{\mathcal A}({\mathbold x}_n,\alpha,\rz)=\{\mu: \text{there exists
  some } \sigma \text{ s.t. }(\mu,\sigma)\in {\mathcal A}({\mathbold
  x}_n,\alpha,\rz\times \rz_+)\}
\end{equation} 
This is equivalent to projecting the approximation region of
Figure~\ref{fig:copper_approx} onto the $x$-axis. The result is the
interval $[1.945,2.087]$. The standard $0.9$ confidence interval for
$\mu$ based on the $t$-statistic is the smaller interval
$[1.978,2.054]$.  If the data really are normally distributed then the
standard confidence interval for $\mu$ will be smaller than the
corresponding approximation interval. If the data are not normally 
distributed then the approximation interval can be smaller, indeed
much smaller than the confidence interval. This will be discussed in
Section~\ref{sec:approx_conf} below.

In (\ref{equ:approx_reg_3}) the same ${\tilde \alpha}$ is used for all
four functionals. There is no need for this. If for example the Kuiper
distance is not regarded as important as the other features it can be
given less weight in terms of a higher value of ${\tilde \alpha}$.

\subsection{Multiple $p$-values}
The approximation region (\ref{equ:approx_reg_3}) is based on the four
statistics $T_i,i=1,\ldots,4$. For each parameter pair $(\mu,\sigma)$
each of the statistics $T_i, i=1,3,4$  comes with a $p$-value
\begin{equation}
p_i(\mu,\sigma)=1-\pr(T_i({\mathbold Y}_n)\le T_i({\mathbold  y}_n)),
\end{equation}
 the statistic $T_2$ comes with the $p$-value
\begin{equation}
p_2(\mu,\sigma)=2\min(\pr(T_2({\mathbold Y}_n)\le T_2({\mathbold
  y}_n)),1-\pr(T_2({\mathbold Y}_n)\le T_2({\mathbold y}_n)))
\end{equation}
where the ${\mathbold Y}_n$ are i.i.d. $N(0,1)$ and ${\mathbold
  y}_n=({\mathbold x}_n-\mu)/\sigma)$. Thus each parameter pair
$(\mu,\sigma)$ comes with four $p$-values attached $p_i(\mu,\sigma),
i=1,\ldots 4$. It belongs to the approximation region if and only
if 
\begin{equation} \label{equ:p_val_gau}
p(\mu,\sigma)=\min(p_i(\mu,\sigma),i=1,\ldots,4) \ge 1-\alpha^*.
\end{equation}
As an example the pair $(2.008,0.110)$ in the approximation region of
Figure~\ref{fig:copper_approx}  has the $p_i$-values
$(0.720,0.683,0.123,0.967)$.

The multiple $p$-values associated with each parameter value stand in
contrast to the usual definition of a $p$-value which uses only one
statistic (see the quotation~\ref{eq:A}).

\subsection{Approximation and confidence regions} \label{sec:approx_conf}
At first sight the  approximation region (\ref{equ:approx_reg_1}) can be
interpreted as a confidence region. If the data ${\mathbold x}_n$ were
indeed generated under some model $P_0 \in {\mathcal P}$ then because
of (\ref{equ:approx_reg_2}) we have
\begin{equation}  \label{equ:approx_reg_4}
\pr(P_0\in {\mathcal A}({\mathbold X}_n(P_0),\alpha,{\mathcal
  P}))=\alpha \text{ for all } P_0 \in {\mathcal P}\,.
\end{equation} 
Such an interpretation however causes difficulties. Consider the family ${\mathcal
  P}=\{N(\mu,\sigma^2): (\mu,\sigma) \in \rz\times\rz_+\}$.  A
standard confidence region for the `true' value $\mu_0$ of $\mu$ is
based on the assumption that there is indeed a `true' value $\mu_0$ of
$\mu$. That is the data were generated under $N(\mu_0,\sigma^2)$ for
some $\sigma$. This assumption is not checked in the formal inference
phase and consequently a confidence region for $\mu_0$ is
never empty. The interpretation is that it is a measure precision with which
$\mu_0$ can be determined. 

\begin{figure}[hb]
\begin{center}
\includegraphics[width=4cm,height=12cm,angle=-90]{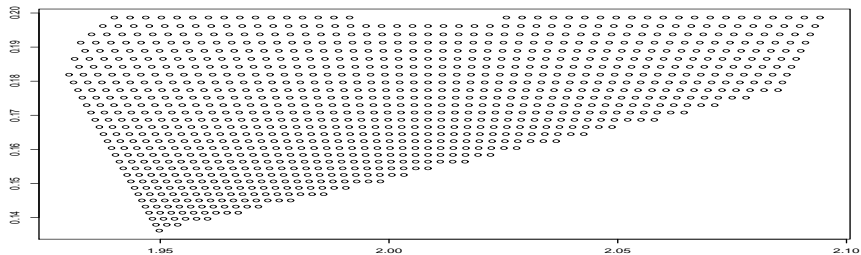}
\end{center}
\caption{The approximation region ${\mathcal A}({\mathbold
  x}_n,0.9,\rz\times \rz_+)$ for the data (\ref{equ:copper_data}) but
with 1.7 replaced by 1.5. \label{fig:copper_approx_2}} 
\end{figure}

The approximation region
(\ref{equ:mu_approx_1}) on the other hand is not based on the
assumption that the data were indeed generated as
i.i.d. $N(\mu,\sigma^2)$ for some $(\mu,\sigma)$. It specifies those
$\mu$-values if any for which $N(\mu,\sigma^2)$ is an adequate
approximation to the data for some $\sigma$. Thus if the adequacy
region (\ref{equ:mu_approx_1}) is small this simply means 
that there are few values of $\mu$ for which $N(\mu,\sigma^2)$ is an
adequate approximation to the data for some $\sigma$. It is not a measure of precision. 
If one imagines the data gradually becoming less and less normal
then the region (\ref{equ:approx_reg_3}) will become smaller and
eventually will be the empty set. One way of doing this is to
gradually increase one value of the sample until this value becomes
incompatible with the feature $T_3$ of (\ref{equ:gauss_func_1}).  
As an example Figure~\ref{fig:copper_approx_2} gives the 0.9
approximation region for the copper data of (\ref{equ:copper_data})
but with the smallest observation of 1.7 being replaced by 1.5
If the 1.7 is replaced by 1.267 the approximation region as calculate
has exactly one point $(1.9812,0.2161)$.

Interpreting (\ref{equ:mu_approx_1}) as a confidence region leads to
complications. As the data become less and less like Gaussian data the
region becomes smaller and smaller  which is interpreted as an
increase in precision. Thus on this interpretation replacing 1.7 by
1.267 in (\ref{equ:copper_data}) leads to exact values for
$(\mu,\sigma)$ namely $(1.9812,0.2161)$. When the region becomes empty
this is as if one goes from infinite precision to no information at
all. A discussion can be found in 
\begin{center}
(*) \quad\href{url}{http://andrewgelman.com/2011/08/25/} .
\end{center}
From the point of view of approximation there is no problem 
of interpretation. The set of adequate parameter values becomes
smaller and smaller and eventually becomes the empty set, that is,
there are no adequate parameter values at all.

\subsection{An empty approximation region} \label{sec:empty_approx}
Consider the approximation region (\ref{equ:approx_reg_3}) with
${\tilde \alpha}=0.975$ corresponding to $\alpha\approx 0.92$. Simulations
show that for normal samples of size 50 the approximation is empty in
about 0.7\% of the cases. This value is based on 5000 simulations. It
is much smaller than the 8\% of the cases where the approximation
region does not contain the $(\mu,\sigma)$ pair used to generate the data.

If the approximation region is empty, that is, the family ${\mathcal
  P}$ contains no model which is an adequate approximation, there may
well be an interest in quantifying just how poor the approximation
is. One way of doing this is to determine the smallest value of
$\alpha$, say $\alpha^*$ such that the approximation region is
non-empty. The corresponding $p$-value is defined as
$p^*=1-\alpha^*$ which is a measure of the goodness of the
approximation: the smaller the $p$-value the worse the approximation.

For the approximation region (\ref{equ:approx_reg_3}) it is always
possible to calculate $p$ as the quantiles $q$ 
can be calculated. If the quantiles were obtained by simulation and
the approximation is poor then it may not be possible to calculate the
$p$-values. An alternative is suggested in \cite{LINLIU09}. It is
based on the idea that it is easier for there to be an adequate
approximation if the sample size is small. Samples ${\mathbold x}_m^*$
of size $m$ are drawn for the original sample ${\mathbold x}_n$ and
the approximation region calculated. The measure of the degree of
approximation is the largest value of $m$ for which the approximation
region is not empty in 50\% of the cases. An example is given in
Chapter 3.8 of \cite{DAV14} for a sample of size $n=189$. The family
of models considered was the family of discretized gamma models and
the concept of adequacy was based on the total variation metric The
fit was so poor that even for $\alpha=0.999999$ there was no adequate
approximation. The size of the smallest random subsamples for which
there was an adequate approximation in 50\% of the cases was
approximately 40.

\subsection{A non-empty approximation region}
If the approximation region is defined by statistics as in
(\ref{equ:approx_reg_3}) then for each model $P$ the $p$-values for each
of inequalities is calculated and the minimum value taken. The maximum
of these values taken over the approximation region is then a measure
of the degree of approximation. Again, the smaller this value the
poorer the approximation. The statistician can base the decision on
whether to use the family of models ${\mathcal P}$  at least in part
on the maximum $p$-value. A cut-off point 0.2 implies that there are adequate
models where the $p$-values of the functionals involved all exceed
0.2. 

\begin{figure}[!]
\begin{center}
\includegraphics[width=4cm,height=12cm,angle=-90]{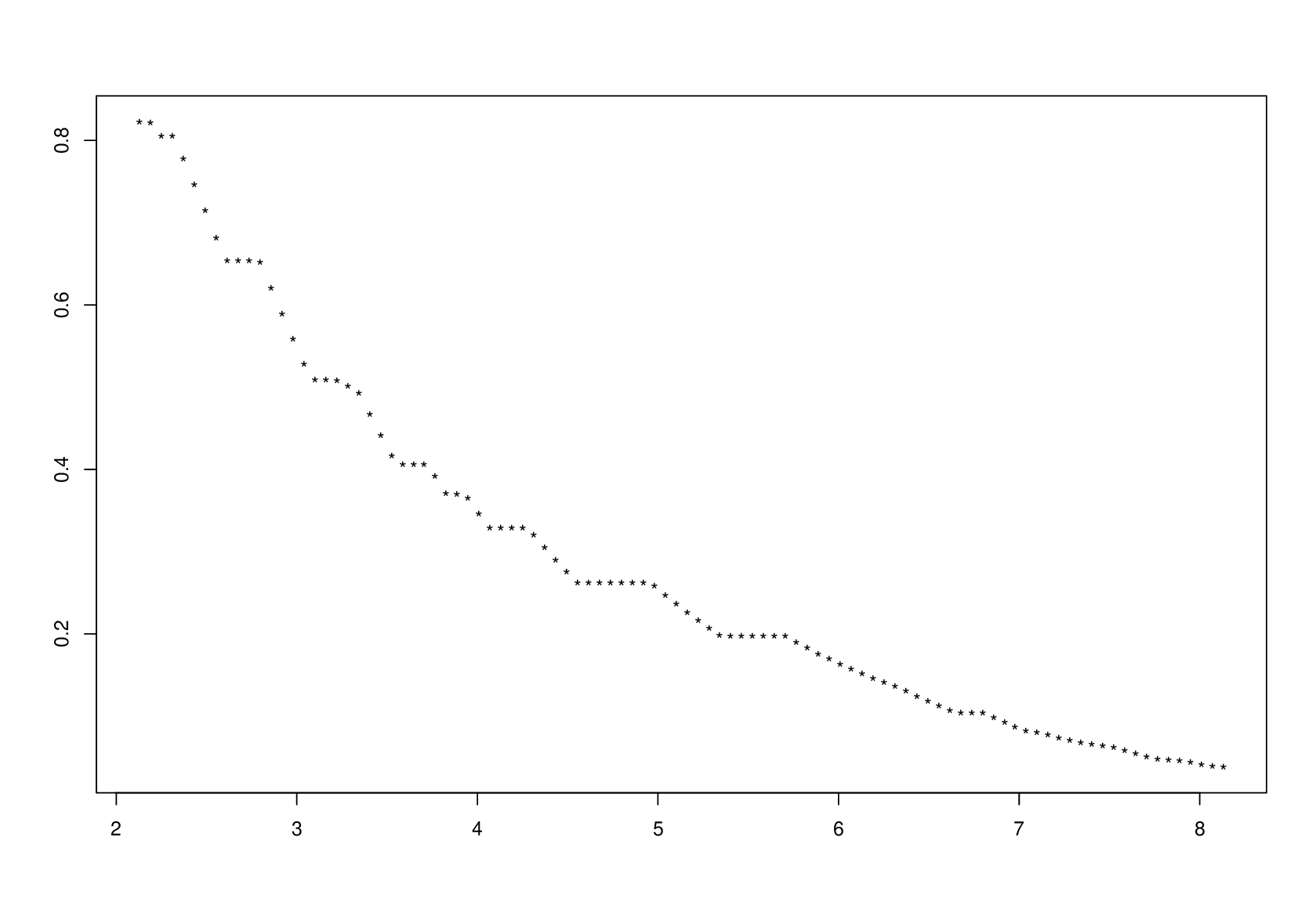}
\includegraphics[width=4cm,height=12cm,angle=-90]{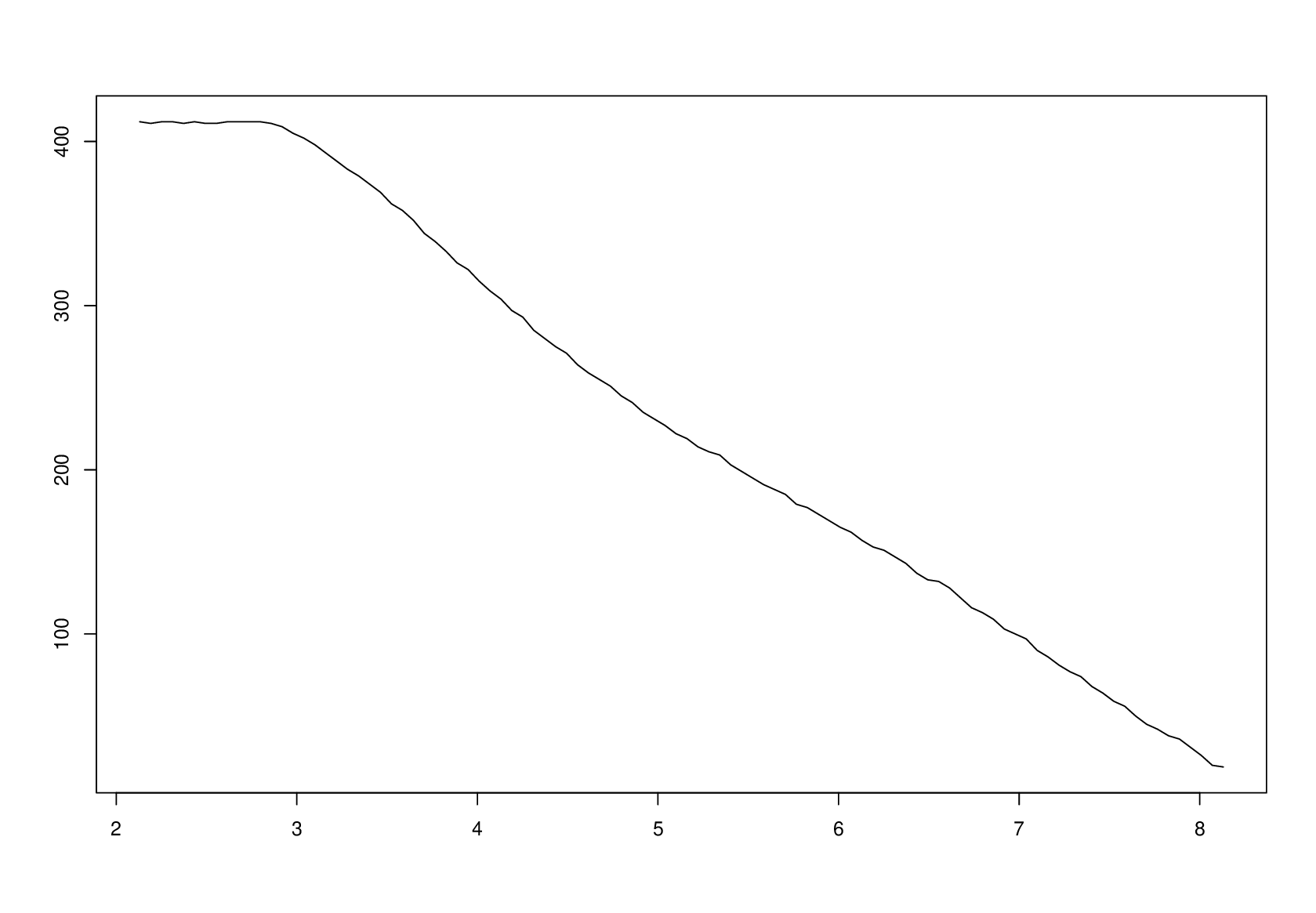}
\end{center}
\caption{Upper panel: the $p$-values  (\ref{equ:p_val_gau}) plotted against the of
  the largest observation for a normal sample of size
  $n=27$. Lower panel: the number of points in the approximation
  region also plotted against the size of the largest
  observation. \label{fig:p_val_out}} 
\end{figure}

Figure~\ref{fig:p_val_out} shows an example of this. A $N(0,1)$ sample
of size $n=50$ was generated and the largest observation 2.130
gradually increased in steps of 0.06875. The upper panel shows the $p$-values
as a function of the size of this observation, the lower panel the
the number of points in the approximation region evaluated over a grid of parameter
values. It can be interpreted as a proxy for the area of the region.
 
The quantiles or the $p$-values of the $p$-values can be obtained from
simulations. For a normal sample of size $n=28$ the 0.001, 0.01, 0.05
and 0.1 quantiles are  0.0713, 0.210, 0.324 and 0.393
respectively. These values are based on 10000 simulations. 

The $p$-value based on the normal family of models is 0.406 with a $p$-value
($p$-value of the $p$-value) of about 0.1. If the smallest observation
1.70 is removed the $p$-value becomes 0.835 which corresponds to
a $p$-value of 0.96. If the smallest value is set to 1.4 the $p$-value
of the $p$-values is about 0.001. This raises the questions to how
bad an approximation can be whilst still basing an analysis on the
family of distributions. 

The following quotation from  \cite{HUB11} is relevant in this
context. It immediately precedes the quotation \ref{eq:D}
\begin{equation}
  \tag{F}\label{eq:F}
  \parbox{\dimexpr\linewidth-6em}{%
    \strut
If a goodness-of-fit tests rejects a model, we are left with many
alternative actions. Perhaps we do nothing (and continue to live with
a model certified to be inaccurate). Perhaps we tinker with a model by
adding or adjusting a few more features (and thereby destroy its
conceptual simplicity). Or we switch to an entirely different model,
maybe one based on a different theory, or maybe in the absence of such
a theory, to a purely phenomenological one. 
In the opposite case, if a goodness-of-fit test does not reject a ...
\strut
  }
\end{equation}

\subsection{$p$-values and hypotheses} \label{sec:hypth_test}
Consider the Gaussian family of models and the null hypothesis
$H_0:\mu\ge \mu_0$. The $p$-value defined by the $t$-statistic
\begin{equation} \label{equ:t_p_val}
\text{pt}\left(\sqrt{n}\,\frac{\text{mean}({\mathbold
     x}_n)-\mu_0}{\text{sd}({\mathbold x}_n)},n-1\right)
\end{equation}
 is often used as a measure as to the extent that $\mu_0$ is
 compatible with the data. This definition is not acceptable from the
 point of view of approximation as it does not specify any value for
 $\sigma$. 

As an example consider the copper data (\ref{equ:copper_data})
Suppose the legal limit is 2.1 milligrams per litre and we wish to
test the hypothesis that this is exceeded. The Gaussian family of
models will be used so that with the usual identification of the amount
of copper in the water with $\mu$ the null hypothesis becomes
\begin{equation} \label{equ:hyp_2.1}
H_0: \mu \ge 2.1.
\end{equation}
The $p$-value as defined by (\ref{equ:t_p_val}) is 0.000436.

An equivalent definition of a standard $p$-value is the
following. Given the  $p$-value $p^*$ put $\alpha^*=1-p^*$. Then
$\alpha^*$ is the smallest value of $\alpha$ such that the confidence
region for $\mu$ contains $\mu_0$. This can be used to define a
$p$-value using the idea of an approximation region. This $p$-value is
defined as $p^*=1-\alpha^*$ where $\alpha^*$ is the smallest value of
$\alpha$ such that the approximation region contains a point
$(\mu_0,\sigma)$ for some $\sigma$. This is similar to the definition
of a $p$-value for an empty approximation region given in
Section~\ref{sec:empty_approx} . If this is done for the water data
(\ref{equ:copper_data}) using the approximation region
(\ref{equ:approx_reg_3}) the resulting $p$-value is 0.045. 

Replace now the smallest value 1.7 by 0.7. The $p$-value of
(\ref{equ:t_p_val}) is now 0.015. At first sight this may seem
surprising as the value 0.7 is less consistent with
(\ref{equ:hyp_2.1}) than is 1.7. The reason is that the standard
deviation is now 0.274 as against 0.116. The $p$-value based on the
approximation region is 0.00018. The value of $\sigma$ is 0.310. The
reason is that the value 0.7 is essentially an outlier. This is picked
up by the statistics $T_3$ and $T_4$ but not by the $t$-statistic. See
the second Huber quotation (\ref{eq:F}). The outlyingness of 0.7 should
have been detected in the EDA phase before moving on to the formal
inference phase. This raises the question of how to react to the outlier. 

\section{$p$-values and functionals}
The purpose of the copper measurements (\ref{equ:copper_data}) it to
give a point estimate of the amount of copper in the sample of
drinking water combined with an interval of reasonable values. The
mean and a confidence interval using a normal model give a reasonable
solution for this particular data sets, but there are
problems. 

One immediate question is why the Gaussian family and not
the Laplace (double exponential) family? This question draws attention
to the fact that the location-scale problem is ill-posed when density
based methods, maximum likelihood or Bayes, are used. Some form of
regularization is required. What are required are `bland' or
`hornless' models (see Section~2, {\it B is   for Blandness}, of
\cite{TUK93B} and Chapter~1.3.6 of \cite{DAV14}). In the
location-scale situation one possible form of regularization is to use
minimum Fisher information models such as the Gaussian. 

Another problem is to relate the parameters of the model to the real
world. As the purpose of the copper data is to estimate
the amount of copper in the water, simply estimating (in another sense
of the word estimate) the parameters of a parametric model does not
solve the problem. The parameters must be connected to the real
world. For the Gaussian family this is not a problem as the canonical
connection is to identify the location parameter $\mu$ with the actual
amount of copper in the water. However this fails for the log-normal
distribution, another minimum Fisher distribution. One can still
associate the actual amount of copper with the mean but also with the
median. This gives two different identifications for the same model.

The final problem is that of outliers. They are common in
interlaboratory tests and any method of analysis must be able to deal
with them. Neither the Gaussian, Laplace or log-normal achieve this. 

The path taken in Chapter~5 of \cite{DAV14} is to use
$M$-functionals (see Chapters~4 and 5 of \cite{HUBRON09}). Given
$\psi$- and $\chi$-functions $\psi$ and $\chi$ respectively and a
probability measure $P$ over $\rz$ the $M$-functional $T_M$ is defined
by $T_M(P)=(T_L(P),T_S(P)$ where $T_L(P)$ and $T_S(P)$ solve
\begin{equation} \label{equ:m-func}
\left\{\begin{array}{ccc}
\int \psi\left(\frac{x-T_L(P))}{T_S(P)}\right)\,dP(x)&=&0\\
\int \chi\left(\frac{x-T_L(P))}{T_S(P)}\right)\,dP(x)&=&0\\
\end{array}
\right.
\end{equation}
The functions $\psi$ and $\chi$ can be so chosen so that (i)
(\ref{equ:m-func}) has a unique solution for all $P$ with a largest
atom of less than 0.5 and (ii) the functional $T_M(P)$ is locally
uniformly Fr\'echet differentiable in a  Kolmogorov neighbourhood of
$P$ see(\cite{DAV98} and page 54 of \cite{HAMRONROUSTA86}). This gives
stability of analysis with respect to  $P$. The functions used here
are
\begin{equation} \label{equ:psi_chi}
\psi(u,c)=\frac{\exp(u/c)-1}{\exp(u/c)+1},\quad
\chi(u)=\frac{u^4-1}{u^4+1}
\end{equation}
where $c$ is a tuning constant set here to 5. 

The connection with reality is achieved by identifying the amount of
copper with $T_L(P)$ for any adequate model $P$. the only form of
adequacy required is that the model $P$ is in a small Kolmogorov
neightbourhood of the empirical distribution  $\ep_n$ of the
data. This still leaves open the choice of $T_M$. This will be
discussed at the end of the section.

Let ${\mathbold X}_n(P)$ denote a sample of size $n$ of i.i.d. random
variable with distribution $P$, and by $\text{q}_{\psi}(\cdot,n,P)$ the
quantiles of 
\[ \frac{1}{\sqrt{n}}\left\vert\sum_{i=1}^n
\psi\left(\frac{X_i(P)-T_L(P)}{T_S(P)}\right)\right\vert \]
with the corresponding definition of $\text{q}_{\chi}(\cdot,n,P)$. The an
$\alpha$-approximation region for the functional $T_M$ is defined by
\begin{eqnarray}
{\mathcal A}({\mathbold x}_n,\alpha,T_M)&=&\Big\{(T_L(P),T_S(P)): d_{ko}(\ep_n,P)<
\text{qdk}(\tilde{\alpha},n),\label{equ:approx_M}\\
&&\quad \sqrt{n}\left\vert
  \int\psi\left(\frac{x-T_L(P)}{T_S(P)}\right)\,d\ep_n(x)\right\vert
\le \text{q}_{\psi}(\tilde{\alpha},n,P),\nonumber\\
&&\quad \sqrt{n}\left\vert 
  \int\chi\left(\frac{x-T_L(P)}{T_S(P)}\right)\,d\ep_n(x)\right\vert
\le \text{q}_{\chi}(\tilde{\alpha},n,P)\Big\} \nonumber
\end{eqnarray}
where $\tilde{\alpha}=(2+\alpha)/3$, $\ep_n$ denotes the empirical
distribution of the data ${\mathbold x}_n$, $d_{ko}$ the Kolmogorov
metric and  $\text{qdk}(\cdot,n)$ its quantile function. The choice of
$\tilde{\alpha}$ corresponds to spending $(1-\alpha)/3$ on each of the
three features in the definition of the approximation region.

As 
\[\ex\left(\psi\left(\frac{X_i(P)-T_L(P)}{T_S(P)}\right)\right)=0\]
and
\[ \ex\left(\psi\left(\frac{X_i(P)-T_L(P)}{T_S(P)}
  \right)^2\right)=\int\psi\left(\frac{x-T_L(P)}{T_S(P)}\right)^2\,dP(x)\] 
it follows from the  central limit theorem that 
\[ \frac{1}{\sqrt{n}}\sum_{i=1}^n
\psi\left(\frac{X_i(P)-T_L(P)}{T_S(P)}\right) \asymp
  N\left(0,\int\psi\left(\frac{x-T_L(P)}{T_S(P)}\right)^2\,dP(x)\right)\,.\]
with the same result for $\chi$. Thus asymptotically
\begin{equation} \label{equ:asymp_quant_m}
\text{q}_{\psi}(\tilde{\alpha},n,P)\approx
\text{qnorm}(\tilde{\alpha})\sqrt{\int\psi
  \left(\frac{x-T_L(P)}{T_S(P)}\right)^2\,dP(x)}
\end{equation}
with the same result for $\chi$. As the random variables are bounded
the normal approximation is good for small values of $n$. 

\begin{figure}[!]
\begin{center}
\includegraphics[width=4cm,height=12cm,angle=-90]{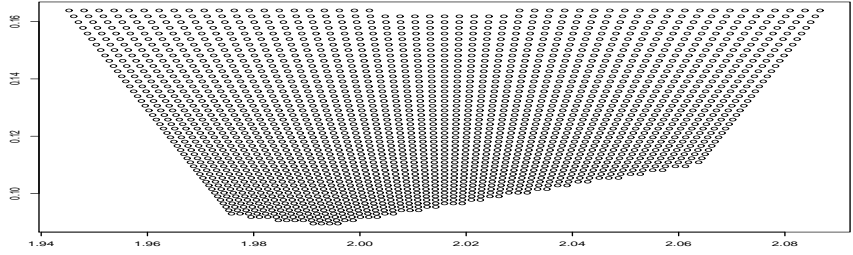}
\end{center}
\caption{The $0.9$  approximation region of the location and scale functionals
  $(T_L,T_S)$ for the copper data using the psi- and chi- functions of
  (\ref{equ:psi_chi}) with $c=5$. \label{fig:gau_m_apprx_reg}}  
\end{figure}

The requirement  $d_{ko}(\ep_n,P)<
\text{qdk}(\tilde{\alpha},n)$ forces $P$ into a $O(1/\sqrt{n})$
Kolmogorov neighbourhood of
$\ep_n$. This together with the locally uniform Fr\'echet
differentiability implies  $\text{q}_{\psi}(\tilde{\alpha},n,P) \approx
\text{q}_{\psi}(\tilde{\alpha},n,\ep_n)$ (see pages 107-108
of\cite{DAV14}) and together with (\ref{equ:asymp_quant_m}) it leads to
the approximate approximation region
\begin{eqnarray}
\lefteqn{\tilde{{\mathcal A}}({\mathbold
    x}_n,\alpha,T_M)=\Big\{(T_L(P),T_S(P)):
  d_{ko}(\ep_n,P)<\text{qdk}(\tilde{\alpha},n)}\label{equ:approx_approx_M}\\
&&\quad \sqrt{n}\left\vert
  \int\psi\left(\frac{x-T_L(P)}{T_S(P)}\right)\,d\ep_n(x)\right\vert
\le \text{qnorm}(\tilde{\alpha})\sqrt{\ev_{\psi}(\ep_n)}\,,\nonumber\\
&&\quad \sqrt{n}\left\vert 
  \int\chi\left(\frac{x-T_L(P)}{T_S(P)}\right)\,d\ep_n(x)\right\vert
\le\text{qnorm}(\tilde{\alpha})\sqrt{\ev_{\chi}(\ep_n)} \nonumber
\end{eqnarray}
where
\[\ev_{\psi}(\ep_n)=\frac{1}{n}\sum_{i=1}^n\psi\left(\frac{x_i-
    T_L(P)}{T_S(P)}\right)^2\,.\]
This approximation to (\ref{equ:approx_M}) can be calculated over a
grid of values. It is shown in Figure~\ref{fig:gau_m_apprx_reg} for
the copper data with $\alpha=0.9$.  It may be compared with the
approximation region based on the Gaussian distribution as shown in
Figure~\ref{fig:copper_approx}.
The approximation region for $T_L(P)$ is obtained by projecting
the approximation region onto the $x$-axis. For the copper data with
$\alpha=0.9$ it is $[1.964,2.067]$ compared with the standard
0.9-confidence interval $[1.978,2.054]$ based on the $t$-statistic.

The approximation region (\ref{fig:gau_m_apprx_reg}) remains unchanged
if the smallest observation 1.7 is replaced by zero. This is in sharp
contrast to the approximation region based on the Gaussian family of
models which is empty in this case. This one example of stability of
analysis deriving from the use of $T_M$: small changes in the data,
here a single data point, lead to only small changes in the result. It
was pointed out above that the location-scale problem requires
regularization. The use of the $M$-functional $T_M$ is a
regularization of the procedure not the models.

Hypothesis testing as in Section~\ref{sec:hypth_test} can be done as
follows. For the copper data the null hypothesis  (\ref{equ:hyp_2.1})
is replaced by 
\[H_0: T_L(P)\ge 2.1.\]
The $p$-value is $p^*=1-\alpha^*$ where $\alpha^*$ is the smallest
value of $\alpha$ such that  (\ref{equ:approx_approx_M}) contains
$(2.1,\sigma)$ for some $\sigma$. Its value is $p^*=0.01$.

The $M$-functional used here is not the only one. There are many
possible choices. Which one to use is an empirical question. A member
of the committee which produced the German DIN standard
(\cite{DIN38402}) for analysing water, waste water and sludge
reported that in his experience the median was better than the mean
but worse than the mean after the elimination of outliers. The final
decision was to use Hampel's redescending $\psi$-function (Example~1
on page 150 of  \cite{HAMRONROUSTA86}) which can be
seen as a smooth version of the mean after eliminating outliers. 

\section{Approximation and prediction} 
\subsection{Prediction}
The concept of adequate approximation can be looked at in terms of
prediction. Given  a number $\alpha$ and based on a model $P$ a
prediction has to be made about a sample ${\mathbold x}_n$. That the
prediction is based on $P$ means that if the sample were generated
under $P$, that is ${\mathbold x}_n={\mathbold X}_n(P)$, then the
prediction would be correct with probability $\alpha$. In making the
prediction is has to be decided which aspects of the data are regarded
as important. In the definition of the approximation region
(\ref{equ:approx_reg_3}) the important aspects are given by the
statistics $T_i, i=1,\ldots,4$.  With $P$ = $N(\mu,\sigma^2)$ the
corresponding prediction is that all the inequalities of
(\ref{equ:approx_inequ_1}) will hold with ${\mathbold y}_n=({\mathbold
  x}_n-\mu)/\sigma$ replacing ${\mathbold Y}_n$. If the prediction is
correct then the model $N(\mu,\sigma^2)$ is accepted as an adequate
approximation to the data.

\subsection{Jeffreys on $p$-values}
The following is often cited as an argument against the use of $p$-values:
\\ 
\begin{equation}
  \tag{G}\label{eq:G}
  \parbox{\dimexpr\linewidth-6em}{%
    \strut
.... gives the probability of departures, measured in a particular
way, equal to or greater than the observed set, and the contribution
from the actual value is nearly always negligible.  What the use of
$P$ implies, therefore, is that a hypothesis that may be true may be
rejected because it has not predicted observable results that have not
occurred. This seems to be a remarkable procedure. On the face of it,
the evidence might more reasonably be taken as evidence for the
hypothesis, not against it.
\strut
  }
\end{equation}
(page 385 of \cite{JEF39}).

Suppose the hypothesis is that the data follow the $N(0,1)$
distribution. What observable results does this hypothesis predict? It
seem pointless to predict a single value as such a prediction would be
wrong with probability 1. The prediction must be a set ${\mathcal S}$
of values with the prediction being regarded as correct if the
observable result $x$  lies in ${\mathcal S}$.  Putting ${\mathcal
  S}=\rz$ results in the prediction being correct with probability one
but this is somewhat vacuous. A non-vacuous prediction can be obtained
by specifying a probability $\alpha$ and a set ${\mathcal S}(\alpha)$
such that the prediction is correct with probability $\alpha$, $\pr(X\in
{\mathcal S}(\alpha))=\alpha$. It is worthy of note that the larger
$\alpha$ the more vacuous the prediction so to speak. As a  simple example put
$\alpha=0.95$ and ${\mathcal S}(\alpha)=(-1.96,1.96)$ and suppose that
$x=3.121$ is observed. The $p$-value is $\pr(\vert X \vert > 3.121)=
0.0018$ and for this to be a successful prediction would require
$\alpha=0.9982$ rather than the chosen $\alpha=0.95$. We now interpret
`not predicted to occur' in the sense `predicted not to
occur' rather than in the sense `forgetting to predict'. If it were
agreed beforehand that a false prediction would lead to the null
hypothesis to be rejected, then this is done because a value predicted
not to occur, namely 3.121, did in fact occur.  This seems an
unremarkable procedure. How bad the prediction error is can be
measured by the $\alpha=0.9982$ required to make the prediction
correct and which corresponds to a very weak prediction in that it
would be correct in 99.8\% of the times.

\section{$p$-values and choice of covariates in stepwise regression} 
The following is based on \cite{DAV16c}. Given a  data set of size $n$
consisting of a dependent variable ${\mathbold y}(n)$ and $p(n)$
covariates ${\mathbold x}(n)$ the problem is to decide which if any of the
covariates to include. The discussion below will be restricted to the
case where $p(n)$ is chosen by stepwise regression but the idea can be
extended to considering all subsets of the covariates as long as
$p(n)$ is not too large, say $p(n)\le 20$ (see \cite{DAV16a}). 

It would seem that all procedures for choosing the covariates are
based on the standard linear model
\begin{equation} \label{equ:standard_model}
{\mathbold Y}(n)={\mathbold X}(n){\mathbold \beta}(n) +{\mathbold \varepsilon}(n).
\end{equation}
The procedure to be described below is not based on this model. The
basic idea is to compare the covariates ${\mathbold x}(n)$ with
covariates which are simply standard Gaussian white noise. A covariate
${\mathbold x}_j$ is included only if it is significantly better than white
noise. 

Suppose that $p_0\le n-2$ with indices ${\mathcal S}_0$ have already been been 
included in the regression and that the sum of squared residuals is
$ss_0$. Denote by $ss_j$ the sum of squared residuals if the covariate
${\mathbold x}_j$ with $j \notin {\mathcal S}_0$ is included. The next
candidate for inclusion is that covariate for which $ss_j$ is
smallest. Including this covariate leads to a sum of squared residuals
\[ ss_{01} =\min_{j \notin {\mathcal S}_0} ss_j.\]
Replace all the covariates not in ${\mathcal S}_0$ in their
entirety by standard Gaussian white noise. Let $SS_j$ denote the sum
of squared residuals if the random covariate corresponding to
${\mathbold x}_j$ is included. The inclusion of the best of the random
covariates leads to a sum of squared residuals
\[ SS_{01}=\min_{j \notin {\mathcal S}_0} SS_j.\]
The probability that the best random covariate is better than the best
of the actual covariates is
\begin{eqnarray*}
\pr(SS_{01} <ss_{01})&=& 1-\pr(SS_{01}  \ge ss_{01})=1-\pr(\min_{j  \notin
  {\mathcal S}_0} SS_j \ge ss_{01})\\
&=&1-\prod_{j \notin {\mathcal S}_0}\pr(SS_j\ge ss_{01})\\
\end{eqnarray*}
It has been shown by Lutz D\"umbgen that
\begin{equation} \label{equ:beta_p}
SS_j\stackrel{D}{=} ss_0(1-B_{1/2,(n-p_0-1)/2})
\end{equation}
where $B_{a,b}$ denotes a beta random variable with parameters $a$ and
$b$ and distribution function $\text{pbeta}(\cdot,a,b)$. From this it
follows that
\[\pr(SS_j\ge ss_{01})=\text{pbeta}(1-ss_{01}/ss_0,1/2,(n-p_0-1)/2)\]
so that finally
\begin{equation} \label{equ:pval_1_exact}
 \pr(SS_{01}\le ss_{01})=
 1-\text{pbeta}(1-ss_{01}/ss_0,1/2,(n-p_0-1)/2)^{p(n)-p_0}.
\end{equation}
This is the $p$-value for the inclusion of the next covariate. The
simplest procedure is to specify $\alpha<1$ and to continue the
stepwise selection until the first $p$-value exceeds $\alpha$. Those
covariates up to but excluding this last one are the selected ones. The
stopping rule is
\begin{equation} \label{equ:stop_rule_1}
ss_{01}> ss_0\left(1-\text{qbeta}((1-\alpha)^{1/(p(n)-p_0)},1/2,(n-p_0-1)/2)\right)
\end{equation}
where $\text{qbeta}(\cdot,a,b)$ is the quantile function of the beta
distribution with parameters $a$ and $b$. 

The procedure is conceptually and algorithmically simple. It requires
no regularization parameter or cross-validation or an estimate of the
error term in (\ref{equ:standard_model}). It is invariant with respect
to affine changes of unit of the covariates and equivariant with
respect to a permutation of the covariates. It can be extended to
non-linear parametric regression, robust regression and the
Kullback-Leibler discrepancy where appropriate. 
 
As an example we take the leukemia data (\cite{GOLETAL99}
\begin{center}
http://www-genome.wi.mit.edu/cancer/\\
\end{center}
which was analysed in \cite{DETBUH03}. These consist of data on $n=72$
samples of tissue with with $p(n)=3571$ covariates. The dependent
variable ${\mathbold y}(n)$ is either 0 or 1 depending on whether the
patient suffers from acute lymphoblastic leukemia or acute myeloid
leukemia. The first five genes in order of inclusion with their
associated $p$-values as defined by (\ref{equ:pval_1_exact}) are as
follows: 
\begin{equation} \label{equ:leukemia_lsq}
\begin{tabular}{cccccc}
gene number&1182&1219& 2888& 1946&2102\\
$p$-value&0.0000&8.57e-4&3.56e-3&2.54e-1&1.48e-1\\
\end{tabular}
\end{equation}
 According to this relevant genes are 1182, 1219 and 2888 and given these
 the remaining 3568 are no better than random noise. This applies to
 the gene 1946 but if a simple linear regression is performed using
 this gene alone its $p$-value in the linear regression is 7.75e-9. This is much
 smaller than the 0.254 in (\ref{equ:leukemia_lsq}). The $p$-value
 (\ref{equ:pval_1_exact}) takes into account the stepwise nature of
 the procedure, in particular that gene 1946 is the best of the
 remaining genes once  the genes 1182, 1219 and 2888 have been included.
A simple linear regression does not take this into account.

The data were gathered in the hope of using the gene expression data
to classify the patients. If the classification is based on  genes 1182, 1219 and
2888. A simple linear regression results in one misclassification.
In \cite{DETBUH03} the authors considered 42 different classification
schemes.  Only two of them resulted in a single
misclassification. They used a 1-nearest-neighbour method based on 25
and 3571 genes. For this particular data set the procedure described
above attains the same result and moreover specifies the relevant
genes.

\bibliographystyle{apalike}
\bibliography{literature}
\end{document}